\documentclass[%
 reprint,
 amsmath,amssymb,
 aps,superscriptaddress
]{revtex4-2}

\usepackage{graphicx}
\usepackage{dcolumn}
\usepackage{bm}
\usepackage{color}

\definecolor{strawberry}{rgb}{1.0,0.0,0.5}

\begin{document}

\preprint{APS/123-QED}

\title{Stability of interlocked self-propelled dumbbell clusters}

\author{Fabian Jan Schwarzendahl}\email{Fabian.Schwarzendahl@hhu.de}
\affiliation{Institut f\"ur Theoretische Physik II: Weiche Materie, Heinrich-Heine-Universit\"at D\"usseldorf, 40225 D\"usseldorf, Germany
}
\author{Abraham Maulean-Amieva}
\affiliation{H. H. Wills Physics Laboratory, University of Bristol, Bristol BS8 1TL, United Kingdom
}%

\author{C. Patrick Royall}
\affiliation{%
Gulliver UMR CNRS 7083, ESPCI Paris, University PSL, 75005 Paris, France
}%

\author{Hartmut L\"owen}%
\affiliation{Institut f\"ur Theoretische Physik II: Weiche Materie, Heinrich-Heine-Universit\"at D\"usseldorf, 40225 D\"usseldorf, Germany
}

\date{\today}

\begin{abstract}
Combining experimental observations of Quincke roller clusters with computer simulations and a  stability analysis, we explore the formation and stability of two interlocked self-propelled dumbbells. For large self-propulsion and significant geometric interlocking, there is a stable joint spinning motion of two dumbbells. The spinning frequency can be tuned by the self-propulsion speed of a single dumbbell, which is controlled by an external electric field for the experiments. For typical experimental parameters the rotating pair is stable with respect to thermal fluctuations but hydrodynamic interactions due to the rolling motion of neighboring dumbbells leads to a break-up of the pair. Our results provide a general insight into the stability of spinning active colloidal molecules which are geometrically locked.
\end{abstract}

\maketitle

\section{Introduction}

Soft matter physics of autonomously driven particles on the micron scale is an expanding research arena which comprises a plethora of synthetic and animate systems ranging from self-propelling colloidal Janus particles to swimming bacteria \cite{marchetti2013hydrodynamics,bechinger2016active,ramaswamy2010mechanics,elgeti2015physics,zottl2016emergent}. Most of the situations occur in an embedding liquid environment giving rise to hydrodynamic interactions between the particles which are induced by the particle motion and transmitted via the solvent flow. When two 
such active particles collide, there can be not only an enhanced particle scattering due to repulsive interactions and self-propulsion \cite{GotzePRE2010,theers2018clustering,guzman2016fission,schwarzendahl2018maximum}, also the inverse effect is possible that the particle self-propulsion leads to an increase in the time for which two particles remain in close contact, i.e. the retention time~\cite{wysocki2015giant}.
During this time, 
particles form transiently a pair of virtually attractive particles. In fact, this is the basic mechanism underlying motility-induced phase separation (MIPS) \cite{cates2015motility,palacci2013living,buttinoni2013dynamical}: an initial transient cluster gives rise to further particle aggregation such that the cluster can grow even to a macroscopic size~\cite{buttinoni2013dynamical,lowen2018active}.

While a cluster of spherical repulsive self-propelled particles is always unstable with respect to fluctuations~\cite{lowen2018active,Zampetaki_private_comm}, this may change for attractions and non-convex geometric particle shapes~\cite{wensink2014controlling,wang2022engineering}. In fact, even passive colloidal particles with nonconvex shapes serve as lock-and-key colloids~\cite{sacanna2010lock,wang2014three,avendano2017packing} establishing strong bonds between repulsive particles if they do interlock. When equipped with activity, such particle composites are ideal building blocks for active colloidal molecules with a dynamical function~\cite{mallory2018active,lowen2018active}.
\begin{figure}
    \centering
    \includegraphics[width=0.5\textwidth]{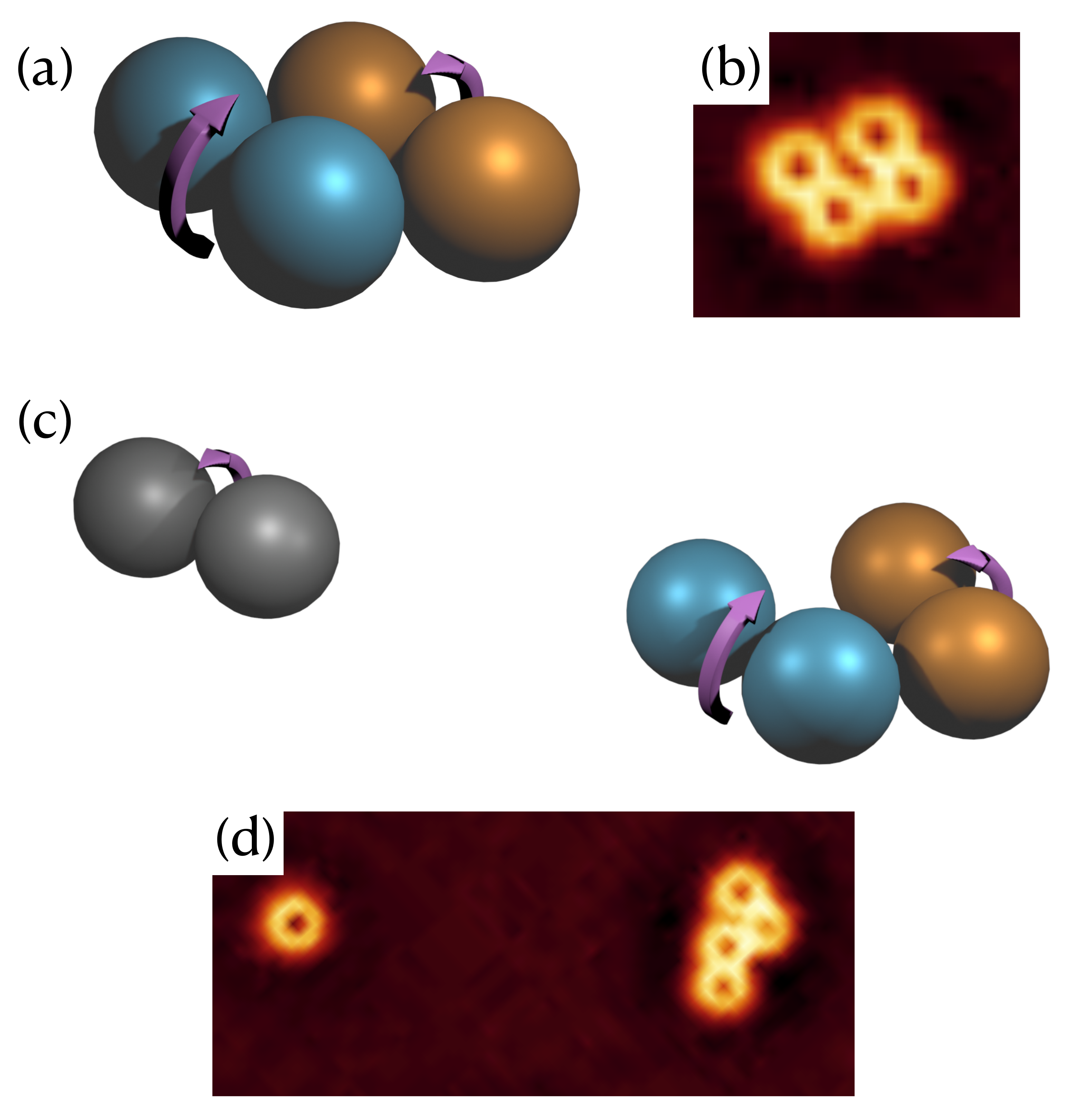}
    \caption{(a) A tetramer of two dumbbells is stable. (b) An experimental tetramer formed by two Quincke rollers. (c) Once a third dumbell passes by the tetramer it breaks up due to long ranged hydrodynamic interactions. (d) A tetramer breaking up in the presence of a third Quincke roller. Experimentally, the sphere diameter is $\sigma=3.1\,\mu\mathrm{m}$.}
    \label{fig:example_dumbbells}
\end{figure}

In this paper, we explore self-propelled particles with a non-convex shape, namely colloidal dumbbells formed by fusing two colloidal spheres together. In contrast to previous studies~\cite{suma2014motility,gonnella2014phase,petrelli2018active,joyeux2016pressure,mandal2017glassy,pirhadi2021dependency,paul2021dynamic}, the direction of self-propulsion is perpendicular to the longer dumbbell axis. When two of these dumbbells collide with opposing self-propulsion directions, they may interlock and stay together for a long time. Recently it was experimentally demonstrated that self-propelled dumbbell Quincke rollers on a two-dimensional substrate form a spinning pair composed of two interlocked dumbbells~\cite{mauleon2021dynamics}. The  spheres of the two spinning dumbells build a tetramer with a parallelogram shape. The spinning frequency can be tuned by the self-propulsion speed.

Here we provide a theoretical analysis for the cluster stability of two interlocked dumbbells and show that in contrast to spheres,  spinning dumbbells pairs are absolutely stable, i.e. they stay together even in the presence of small external fluctuations. The stability, however, depends on the interaction between the individual spherical monomers forming the different dumbbells. If the potential energy between two monomers strongly increases 
as a function of monomer separation, the interaction is harsh and the dumbbells are interlocked almost in a geometric way as two touching  nonconvex hard bodies. In this extreme case there are excluded volume interactions as realized for sterically stabilized colloids and  the experiments of Ref.~\cite{mauleon2021dynamics}. Under these condtions, 
stability occurs for a broad range of intermediate self-propulsion velocities. Clearly for very small self-propulsion, we are close to the passive case where stability is lost. In the opposite limit, extremely large self-propulsions  dominate any potential energy barrier leading ultimatively to an instability.  However, for soft monomer interactions, as realized for charged dumbbells~\cite{knapek2022compact,nosenko2020active,arkar2021dynamics} or dumbbells with soft polymeric shells~\cite{likos1998star} or composed of microgel particles~\cite{alvarez2021reconfigurable}, the degree of interlocking is much weaker. In this case we show that the tetramer is always unstable.  
We further explore the stability for geometric interlocking (as relevant for the experiment~\cite{mauleon2021dynamics}) against thermal fluctuations and show by computer simulations that these fluctuations are irrelevant for the experimental determined life-times of the spinning dumbbell pairs. In other words, thermal fluctuations are too weak to explain the observed finite duration of the spinning cluster. However, long-ranged hydrodynamic interactions between a spinning tetramer and a neigbouring dumbbell can destroy the spinning pair much more efficiently. This is demonstrated by experimental snapshots and supported by a computer simulation study of Quincke rollers which includes hydrodynamic interactions between the particles and the underlying substrate. For the latter, the multiparticle collision dynamics method is employed~\cite{gompperBookChap2009}.

The implications of our work are manifold: first, it is important to understand stability of spinning active colloidal molecules~\cite{schmidt2019light} both from a fundamental point of view~\cite{soto2014self,soto2015self} and for applications such as micromixers~\cite{cai2017review}. Second,  stable spinning colloidal molecules can used as building blocks of rotating particles forming a chiral fluid which is a booming area of present research~\cite{nguyen2014emergent,banerjee2017odd,han2021fluctuating,scholz2021surfactants,poggioli2022odd,zhang2022hyperuniform}. Third, our work provides an analytical framework to study more complicated non-convex colloidal shapes and their impact on cluster stability as basic building blocks for more complex colloidal molecules with a dynamical function~\cite{mallory2018active}.

This paper is organized as follows: First in Sec.~\ref{Sec:theory}, the stability of tetramers is studied using a linear stability analysis and the tetramers spinning frequency is compared to experiments. Then, in Sec.~\ref{Sec:exp} the experimental approach is discussed. In Sec.~\ref{Sec:BD} Brownian dynamics simulations are used to assess the impact of fluctuations on the stability of tetramers. Finally, in Sec.~\ref{Sec:hydro} the influence of hydrodynamic interactions on the stability are studied and the break-up of tetramers by a third colloid passing by is investigated.

\section{Theory}
\label{Sec:theory}
\begin{figure*}
    \centering
    \includegraphics[width=0.7\textwidth]{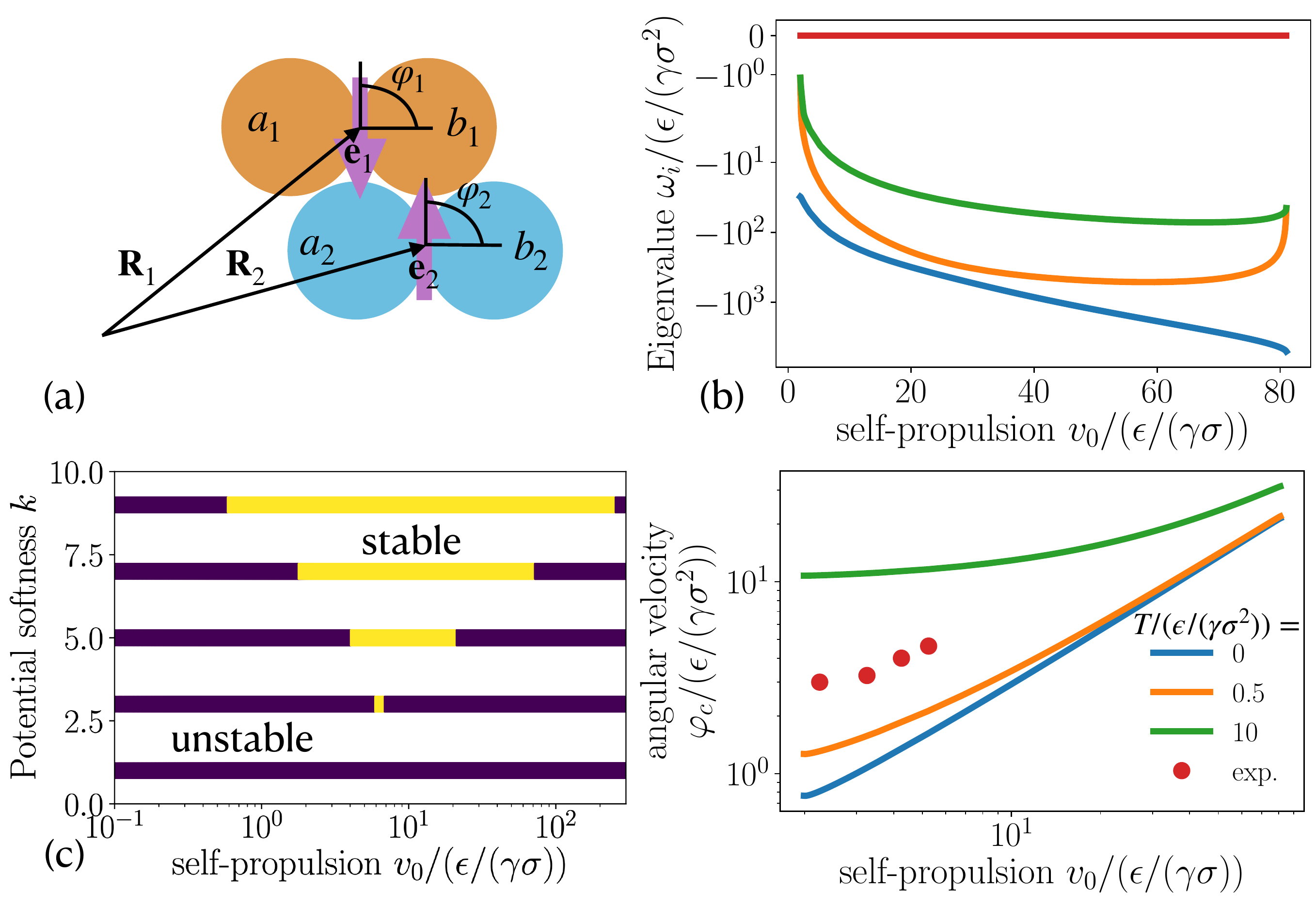}
    \caption{(a) Sketch of our theoretical setup with two jammed dumbbells (denoted by $1$, and $2$) with their respective circles (denoted by $a$ and $b$) and their orientations ($\bm{e}_1$ and $\bm{e}_2$). (b) Eigenvalues as a function of self-propulsion velocity $v_0$ (here using the softness $k=9$ for the interparticle potential). (c) Stability diagram showing regions with a stable tetramer (yellow) and regions which are unstable (blue) for varying self-propulsion velocity $v_0$ and potential softness $k$. (d) Angular velocity of the tetramer as a function of self-propulsion velocity for experiments (data was extracted from~\cite{mauleon2021dynamics}) and for theory with different $T$.}
    \label{fig:fig2}
\end{figure*}
The stability of two interlocked dumbbells is first analyzed theoretically, by using a two dimensional projection of the system. The dumbbells are modelled as active rigid particles, that self propel along their short axis (see Fig.~\ref{fig:fig2}(a)) and are moving in the two-dimensional $xy$-plane. Explicitly, the following overdamped equations of motion for the two dumbbells are used
\begin{align}
    \frac{\mathrm{d}\bm{R}_1}{\mathrm{d}t}
    = v_0 \bm{e}_1 
    + \frac{1}{\gamma} (
    \bm{F}_{a_2 \rightarrow a_1}
    +\bm{F}_{a_2 \rightarrow b_1}
    +\bm{F}_{b_2 \rightarrow b_1}
    +\bm{F}_{b_2 \rightarrow a_1}),
    \label{eq:COM1pos}
    \\
    \frac{\mathrm{d}\bm{R}_2}{\mathrm{d}t}
    = v_0 \bm{e}_2
    + \frac{1}{\gamma} (
    \bm{F}_{a_1 \rightarrow a_2}
    +\bm{F}_{b_1 \rightarrow a_2}
    +\bm{F}_{b_1 \rightarrow b_2}
    +\bm{F}_{a_1 \rightarrow b_2}),
        \label{eq:COM2pos}
\end{align}
where $\bm{R}_1$ is the first and $\bm{R}_2$ the 
center of mass position of the second dumbbell. The dumbbells propel with a constant velocity $v_0$ along their respective axis $\bm{e}_1 = (\mathrm{cos}(\varphi_1),\mathrm{sin}(\varphi_1))^T$ and $\bm{e}_2=(\mathrm{cos}(\varphi_2),\mathrm{sin}(\varphi_2))^T$, where $\varphi_1$ and $\varphi_2$ are the angles with respect to the $x$-axis in a Cartesian coordinate frame and $()^T$ denotes the matrix transpose. The forces of the circle $i$ acting on circle $j$ are denoted by $\bm{F}_{i \rightarrow j}$ where $i$ and $j$ run through all respective circles (see Fig.~\ref{fig:fig2}(a), $i,j \in \{a_1,a_2,b_1,b_2\}$) and $\gamma$ is the friction coefficient.  
The orientational dynamics taking into account the torques on the center of mass positions of the particles are given by
\begin{align}
    \frac{\mathrm{d}\varphi_1}{\mathrm{d}t}
    = \frac{r_0}{\gamma_r} \left[ 
    \bm{e}_1^{\perp}
    \times (
    \bm{F}_{b_2 \rightarrow b_1}
    +\bm{F}_{a_2 \rightarrow b_1}
    )\right.
    \\
   - \left.\bm{e}_1^{\perp}
    \times (
    \bm{F}_{a_2 \rightarrow a_1}
    +\bm{F}_{b_2 \rightarrow a_1}
    ) 
    \right] + T,\nonumber
    \\
      \frac{\mathrm{d}\varphi_2}{\mathrm{d}t}
    = \frac{r_0}{\gamma_r} \left[ 
    \bm{e}_2^{\perp}
    \times (
    \bm{F}_{a_1 \rightarrow a_2}
    +\bm{F}_{b_1 \rightarrow a_2} 
    )\right.
    \\
   - \left.\bm{e}_1^{\perp}
    \times (
    \bm{F}_{b_1 \rightarrow b_2}
    +\bm{F}_{a_1 \rightarrow b_2} 
    ) 
    \right] + T,\nonumber  
\end{align}
where $\bm{e}_1^{\perp}=(-\mathrm{sin}(\varphi_1),\mathrm{cos}(\varphi_1))^T$ is the vector perpendicular to $\bm{e}_1$ and $\bm{e}_2^{\perp}=(-\mathrm{sin}(\varphi_2),\mathrm{cos}(\varphi_2))^T$ the vector perpendicular to $\bm{e}_2$. Here, $r_0$ is the radius of a monomer 
and $\gamma_r$ is the rotational friction coefficient. The constant $T$ stems from a torque, which may arise in an experiment from a slight asymmetry of the two circles, giving rise to constant rotation of a dumbbell. The two dimensional cross product is defined here as $\bm{a} \times \bm{b}= a_x b_y - a_y b_x$ where $a_x$ is the first component of a vector $\bm{a}$ and $a_y$ is the second component.

As a first step of the stability analysis a configuration of the dumbbells needs to be found where the forces on each particle are zero, explicitly 
\begin{align}   
    0
    = v_0 \bm{e}_1 
    + \frac{1}{\gamma} (
    \bm{F}_{a_2 \rightarrow a_1}
    +\bm{F}_{a_2 \rightarrow b_1}
    +\bm{F}_{b_2 \rightarrow b_1}
    +\bm{F}_{b_2 \rightarrow a_1}),
    \label{eq:zeroforce1}
    \\
    0
    = v_0 \bm{e}_2
    + \frac{1}{\gamma} (
    \bm{F}_{a_1 \rightarrow a_2}
    +\bm{F}_{b_1 \rightarrow a_2}
    +\bm{F}_{b_1 \rightarrow b_2}
    +\bm{F}_{a_1 \rightarrow b_2}).
        \label{eq:zeroforce2}
\end{align}
These equations are solved for $\bm{R}_1$ and $\bm{R}_2$ (note that the forces $F_{i\rightarrow j}$ are functions of the positions) to find positions $\bm{R}_{1,s}$ and $\bm{R}_{2,s}$ with no net force acting on the dumbbells. Without restriction of generality the directions displayed in Fig.~\ref{fig:fig2}(a) with angles $\varphi_{1}=\varphi_{1,s}=\frac{3}{2} \pi$ and $\varphi_{2}=\varphi_{2,s}=\frac{1}{2} \pi$ are used to solve Eqs.~\eqref{eq:zeroforce1}-\eqref{eq:zeroforce2}. Further, $\bm{R}_{2,s}=0$ can be used because of translational invariance of the system such that only $\bm{R}_{1,s}$ needs to be found. Typically, there are several solutions for $\bm{R}_{1,s}$ but only one corresponds to an interlocked dumbbell as in Fig.~\ref{fig:fig2}(a). This tetramer solution also yields a condition on $v_0$.

We continue by making a linear stability analysis around the zero force state. 
By using the coordinates
\begin{align}
    \bm{R}_{\mathrm{COM}} = \frac{1}{2}(\bm{R}_1+\bm{R}_2),
    \\
    \bm{R}_{\mathrm{REL}} = \frac{1}{2}(\bm{R}_1-\bm{R}_2),
\end{align}
for the tetramers center of mass $\bm{R}_{\mathrm{COM}}$ and the relative coordinate of the dumbbells $\bm{R}_{\mathrm{REL}}$ the 
Eqs.\eqref{eq:COM1pos}-\eqref{eq:COM2pos} become
\begin{align}
    \frac{\mathrm{d} \bm{R}_{\mathrm{COM}}}{\mathrm{d}t}
    &= \frac{1}{2} v_0 (\bm{e}_1 +\bm{e}_2),
    \\    
    \frac{\mathrm{d} \bm{R}_{\mathrm{REL}}}{\mathrm{d}t}
    &=\frac{1}{2} v_0 (\bm{e}_1 -\bm{e}_2)
    \\
    &+ \frac{1}{\gamma} (
    \bm{F}_{a_2 \rightarrow a_1}
    +\bm{F}_{a_2 \rightarrow b_1}
    +\bm{F}_{b_2 \rightarrow b_1}
    +\bm{F}_{b_2 \rightarrow a_1}) \nonumber.
\end{align}
Here, only the equation for $\bm{R}_{\mathrm{REL}}$ needs to be taken into account in the stability analysis, since $\bm{R}_{\mathrm{COM}}$ is stable due to translational symmetry.
Next, the relative coordinate $\bm{R}_{\mathrm{REL}}$ is decomposed into polar coordinates
\begin{align}
    \bm{R}_{\mathrm{REL}}= R \begin{pmatrix}\mathrm{cos}(\theta) \\ \mathrm{sin}(\theta)\end{pmatrix}.
\end{align}
This results in 4 equations (one for each coordinate $R$, $\theta$, $\varphi_1$ and $\varphi_2$) that are expanded around the zero force state. The result is a $4\times4$ matrix whose eigenvalues $\omega_i$ ($i \in \{1,2,3,4\}$) give the stability of the system, which we solve using computer algebra. 

For analytical tractability, the forces between the dumbbells circles are modeled by a repulsive inverse power law potential
\begin{align}
    \Phi_{i \rightarrow j}= \epsilon \frac{\sigma^k}{|\bm{r}_i- \bm{r}_j|^k},
\end{align}
where the exponent $k$ determines the softness of the interaction, $\bm{r}_i$ is the position of circle $i$ and $\bm{r}_j$ is the position of circle $j$. Here, $\sigma= 2r_0$ is the diameter of a circle and $\epsilon$ is the energy scale. The forces are then calculated using $\bm{F}_{i \rightarrow j}= -\nabla_i \Phi_{i \rightarrow j}$.

Figure~\ref{fig:fig2}(b) shows the four eigenvalues resulting from the stability analysis with a potential softness $k=9$. The interval for $v_0$ that is shown is the range in which there is a tetramer solution to the zero force condition Eqs.~\eqref{eq:zeroforce1}-\eqref{eq:zeroforce2} resembling Fig.~\ref{fig:fig2}(a). Within this region, there are three eigenvalues which are negative and one eigenvalue that is zero. The zero mode gives rise to a constant rotation of the dumbbells around each other, while the negative eigenvalues imply that the tetramer is stable. Outside of the region shown for $v_0$ in Fig.~\ref{fig:fig2}(b) there is no hexamer solution of Eqs.~\eqref{eq:zeroforce1}-\eqref{eq:zeroforce2}. Further, the stability analysis for all other solutions to Eqs.~\eqref{eq:zeroforce1}-\eqref{eq:zeroforce2} yields at least one positive eigenvalue and therefore these solutions are unstable.

The softness of the repulsive potential is controlled by the exponent $k$, which has a strong impact on the stability of the tetramer. For an exponent $k=1$ the system is always unstable, i.e. there is at least one positive eigenvalue (see Fig.~\ref{fig:fig2}(c)), while increasing the exponent leads to a larger stable region (Fig.~\ref{fig:fig2}(c)). In fact, within the region where a tetramer configuration is found using Eqs.~\eqref{eq:zeroforce1}-\eqref{eq:zeroforce2}, the tetramer is always stable. Conversely, once no tretramer solution of Eqs.~\eqref{eq:zeroforce1}-\eqref{eq:zeroforce2} that is akin to Fig.~\ref{fig:fig2}(a) is found, the system is unstable, since it has at least one positive eigenvalue. The stability of the tetramer is not affected by the constant rotation mediated by $T$. 

The spinning frequency $\varphi_c$ of a tetramer has two contributions: the constant rotation mediated by $T$ and an emergent rotation due to the dumbbells active motion. The tetramers spinning frequency has also been measured in ~\cite{mauleon2021dynamics}. To compare these experimental measurements of the spinning frequency, the energy scale $\epsilon$ has the be estimated. Assuming that the force exerted by colliding colloids balances the steric forces obtained from the colloids potential gives the estimate $\epsilon = v \gamma \sigma$. In the experiments a typical velocity at which particles start jamming is $v \approx 40 \sigma/$s (estimated from~\cite{mauleon2021dynamics}), the friction coefficient is estimated by $\gamma= 3 \pi \eta \sigma$, with the viscosity of water $\eta \approx 10^{-3}$ Pas and the diameter of a colloid $\sigma \approx 3 \mu$m. This yields an energy scale of $\epsilon \approx 2.7 \times10^3 k_b$T. The spinning frequency of a single dumbbell was experimentally measured as $T= 20/$s$ \approx 0.5 \epsilon/(\gamma \sigma^2)$ (estimated from~\cite{mauleon2021dynamics}).
Figure~\ref{fig:fig2}(d) shows the experimentally measured spinning frequency of tetramers (data from~\cite{mauleon2021dynamics}) compared to the theoretical prediction (orange line), which match qualitatively and lie in the right order of magnitude without any fitting parameters. 

\section{Experiments and Simulations}

\subsection{Experiments}
\label{Sec:exp}

We prepare the colloidal dumbbells as follows.
We use polystyrene beads (Fluoro-Max, Thermo Fischer) of size $\sigma = 3.1 \mu$m, and polydispersity $5 \%$. The initial suspension is aqueous. Colloids are repeatedly washed with a 0.15 M solution of AOT surfactant in hexadecane. In the absence of a stabilising layer, colloidal clusters form due to van der Waals attractions. We obtain a mixture of clusters as the aqueous solvent is replaced by the low polar solution. Centrifugation is used to separate small dumbbells from the rest of the suspension.

For imaging, a dilute mixture of clusters is loaded into a sample cell fabricated with conductive ITO-coated glass slides. Two slides are separated by a $30 \mu$m-thickness spacer made of optical glue and larger beads. An dc electric field $E$ is applied to the suspension to observe the \emph{Quincke} rotation of colloids \cite{bricard2013emergence}. Image sequences are obtained at 660 fps using bright-field microscopy.
Further details of the experiments may be found in \cite{mauleon2021dynamics}.

\subsection{Brownian dynamics}
\label{Sec:BD}
\begin{figure}
    \centering
    \includegraphics[width=0.5\textwidth]{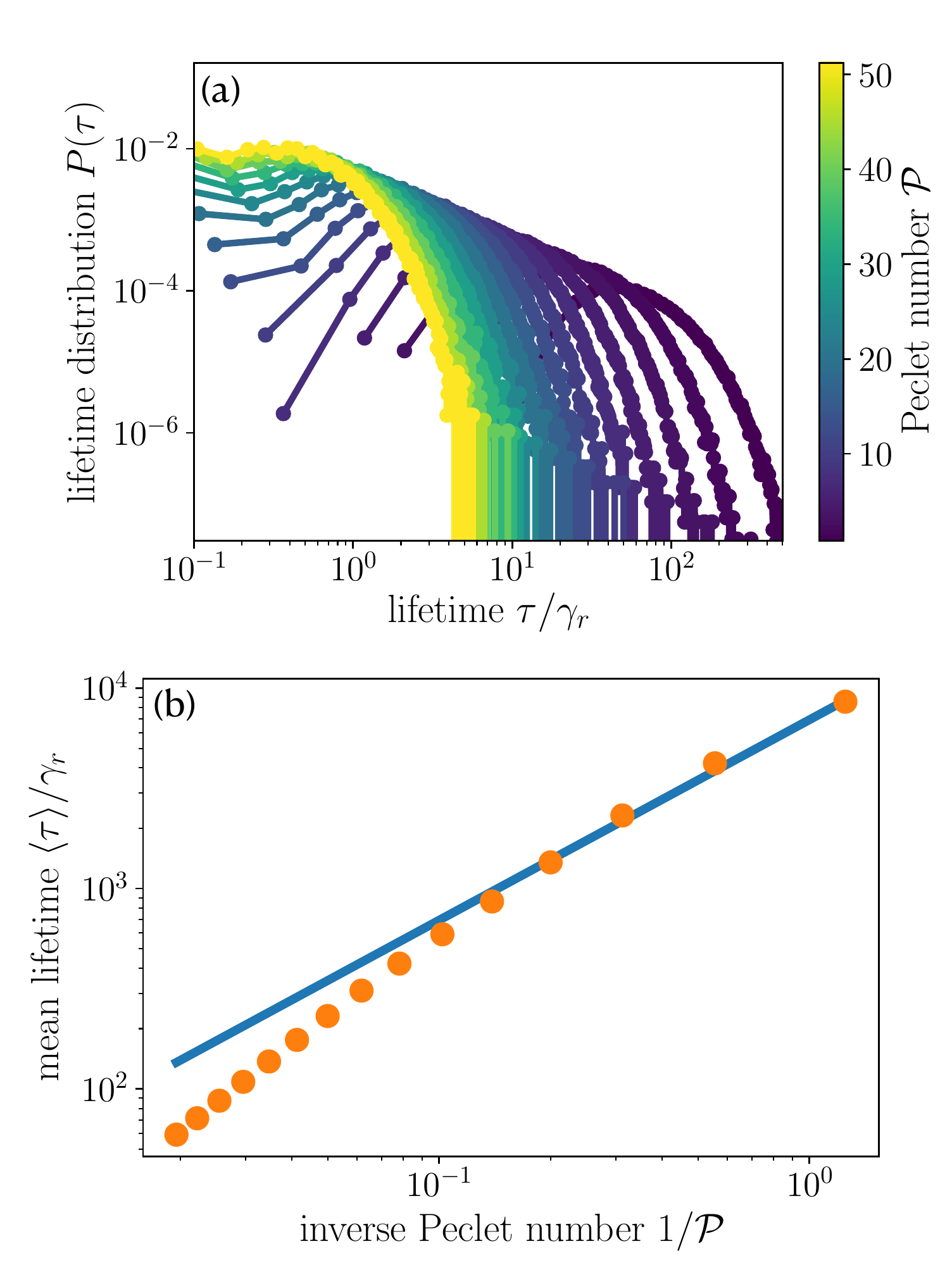}
    \caption{(a) Lifetime distribution $P(\tau)$ for different P\'eclet numbers (color code). (b) Mean lifetime as a function of inverse P\'eclet number (orange circles). Blue line shows a linear fit to the mean lifetimes, which is used to extrapolate the lifetime in an experiment with Quincke rollers ($\approx 10^3$ days).}
    \label{fig:fig3}
\end{figure}
To incorporate
thermal fluctuations, two dimensional Brownian dynamics simulations are employed, with the following equations of motion for the dumbbell's center of mass positions 
\begin{align}
    \frac{\mathrm{d}\bm{R}_1}{\mathrm{d}t}
    = v_0 \bm{e}_1 
    + \frac{1}{\gamma} \bm{F}_{\mathrm{tot},1} + \sqrt{2 D^t_{\parallel}} \eta_{1 \parallel} \bm{e}_1 
    + \sqrt{2 D^t_{\perp}} \eta_{1 \perp} \bm{e}_1^{\perp},
    \\
    \frac{\mathrm{d}\bm{R}_2}{\mathrm{d}t}
    = v_0 \bm{e}_2
    + \frac{1}{\gamma} \bm{F}_{\mathrm{tot},2}
    + \sqrt{2 D^t_{\parallel}} \eta_{2 \parallel} \bm{e}_2 
    + \sqrt{2 D^t_{\perp}} \eta_{2 \perp} \bm{e}_2^{\perp}.
\end{align}
Here, the total forces are given by $
 \bm{F}_{\mathrm{tot},1}=\bm{F}_{a_2 \rightarrow a_1}
    +\bm{F}_{a_2 \rightarrow b_1}
    +\bm{F}_{b_2 \rightarrow b_1}
    +\bm{F}_{b_2 \rightarrow a_1}$ and  $\bm{F}_{\mathrm{tot},2}= 
    \bm{F}_{a_1 \rightarrow a_2}
    +\bm{F}_{b_1 \rightarrow a_2}
    +\bm{F}_{b_1 \rightarrow b_2}
    +\bm{F}_{a_1 \rightarrow b_2}$ for dumbbell $1$ and $2$ respectively. Further, fluctuations with zero mean and correlator $\langle \eta_{i}(t) \eta_{j}(t') \rangle =2 \delta_{ij} \delta(t-t')$ with $i,j \in \{1\parallel, 1\perp, 2\parallel, 2\perp\}$ are included. The fluctuations along the orientations $\bm{e}_i$ have a translational diffusion coefficient $D^t_{\parallel}$ and perpendicular the orientation the diffusion coefficient is $D^t_{\perp}$.
The orientational dynamics of the two dumbbells are given by
\begin{align}
    \frac{\mathrm{d}\varphi_1}{\mathrm{d}t}
    = \frac{r_0}{\gamma_r} 
    T_1 + \sqrt{2 D^r} \xi_1,
    \\
      \frac{\mathrm{d}\varphi_2}{\mathrm{d}t}
    = \frac{r_0}{\gamma_r}  T_2 + \sqrt{2 D^r} \xi_2,
\end{align}
where $T_1 =
    \bm{e}_1^{\perp}
    \times (
    \bm{F}_{b_2 \rightarrow b_1}
    +\bm{F}_{a_2 \rightarrow b_1}
    )
   -\bm{e}_1^{\perp}
    \times (
    \bm{F}_{a_2 \rightarrow a_1}
    +\bm{F}_{b_2 \rightarrow a_1}
    ) $ and $T_2=
    \bm{e}_2^{\perp}
    \times (
    \bm{F}_{a_1 \rightarrow a_2}
    +\bm{F}_{b_1 \rightarrow a_2}
    )
   - \bm{e}_1^{\perp}
    \times (
    \bm{F}_{b_1 \rightarrow b_2}
    +\bm{F}_{a_1 \rightarrow b_2}
    )$ summarize the effects of torques acting on the dumbbells. Further, fluctuations are included with zero mean, correlations $\langle \xi_{i}(t) \xi_{j}(t') \rangle =2 \delta_{ij} \delta(t-t')$ and the rotational diffusion coefficient $D^r$.
    
The forces between the dumbbell's circles are calculated using a Weeks--Chandler--Andersen (WCA) potential \cite{WeeksJCP1971}, which is a common model potential for colloidal particles. Note that in the theoretical treatment a repulsive power law potential was used for analytical tractability, while in the simulations a more realistic WCA potential can be used. Explicitly, the potential used in the simulations reads
\begin{equation} \label{eq:WCAPotentialBD}
  \Phi (r_{ij}) = 
            4 \epsilon \left[ \left(\frac{\sigma}{r_{ij}} \right)^{12}-  
                              \left(\frac{\sigma}{r_{ij}} \right)^6 \right] 
            + \epsilon,          
\end{equation}
if $r_{ij} < 2^{1/6} \sigma$, and $\Phi (r_{ij}) = 0$ otherwise, where $\sigma= 2 r_0$ is the diameter of a dumbbell's circle and $\epsilon=k_B T$ is the energy scale, which is connected to the translational diffusion coefficient of a single circle by $D^t= k_B T/ \gamma$.

To simplify the parameter space the diffusion coefficients $D^t_{\parallel}= \lambda_{\parallel} D^t$, $D^t_{\perp}= \lambda_{\perp} D^t$ and $D^r= 3 D^t/(4 r_0^2)$ are used, where last relation stems from the translational and rotational friction coefficient of a sphere. By that simplification the system is characterized by a single dimensionless number, which is the P\'eclet number $\mathcal{P}= 2 r_0v_0/D^t$. The parameters $ \lambda_{\parallel} = 0.783$ and $\lambda_{\perp} =0.703$ were explicitly measured in experiments.

The simulations are started with a stable configuration as depicted in Fig.~\ref{fig:fig2}(a). Each simulation is run until the center of mass positions ($\bm{R}_1$ and $\bm{R}_2$) of the two dumbbells are separated from each other by a distance $d_{\mathrm{sep}}= 3 \sigma$. While the dumbbells are in contact 
they perform a rotation around each other as theoretically predicted. The time at which $d_{\mathrm{sep}}$ is reached is defined as the lifetime $\tau$ of a tetramer. The distribution of tetramer lifetimes is shown in Fig.~\ref{fig:fig3}(a) for different P\'eclet numbers. Irrespective of the P\'eclet number the lifetime distribution first shows a maximum and then drops with an exponential tail. Clearly, low P\'eclet numbers show (on average) a large tetramer lifetime, while large P\'eclet numbers yield shorter tetramer lifetimes. This trend is also observed in the mean lifetime $\langle \tau \rangle$ (see Fig.\ref{fig:fig3}(b)). 
\begin{figure*}
    \centering
    \includegraphics[width=0.8\textwidth]{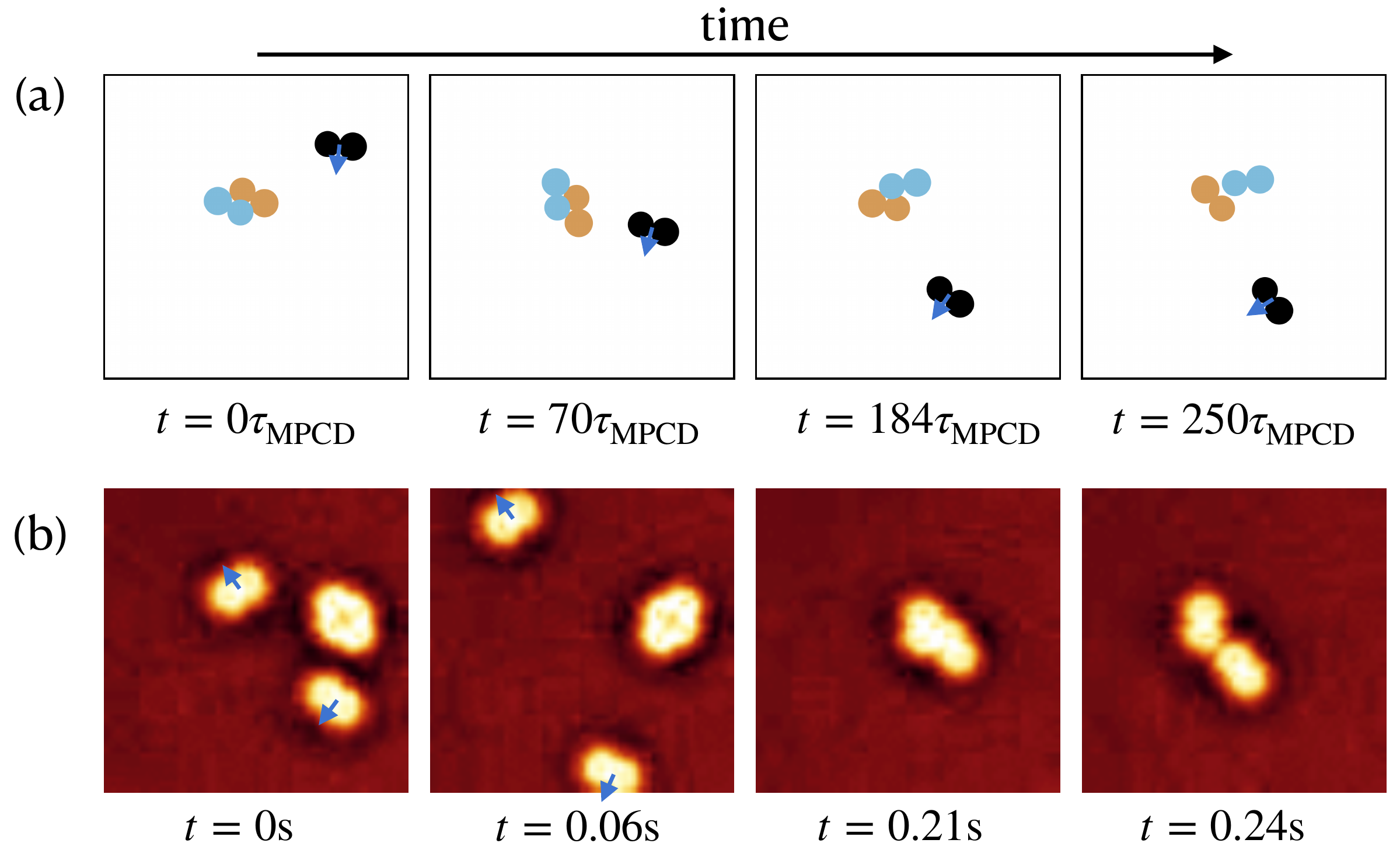}
    \caption{(a) Top view of the hybrid MPCD-MD simulation of rollers where a tetramer (combination of blue and orange dumbbell) breaks up due to the long ranged hydrodynamic interaction of the third dumbbell passing by. (b) Experiments with Quincke rollers showing the breaking of a tetramer after other Quincke rollers passed by (blue arrows indicate the dumbell's direction of motion). Here the diameter of the monomers is $3.0$ $\mu$m.}
    \label{fig:fig4}
\end{figure*}
Extrapolating from Fig.~\ref{fig:fig3}(b) gives an estimate of the expected lifetime for Quincke rollers in an experimental setting. The experimentally measured self propulsion velocity is $v_0 = 200 \sigma/$s. 
Using the translational diffusion coefficient $D^t= \frac{k_B T}{3 \pi \eta \sigma} $, with room temperature $T= 293$K and the viscosity of water $\eta= 1$mPas, the P\'eclet number of a Quincke roller dumbbell is $\mathcal{P} \approx 10^4$. Using a linear extrapolation from the simulation data in Fig.~\ref{fig:fig3}(b) then yields a lifetime of $\langle \tau \rangle \approx 10^3$ days, which is orders of magnitude larger than the experimentally measured lifetimes. Hence, thermal fluctuations alone are not enough to explain the breaking up of tetramers in an experimental setting. 

\subsection{Hydrodynamic interactions}
\label{Sec:hydro}
In order to asses the effect of hydrodynamic interactions between the rollers, we employ a hybrid simulation technique using multiparticle collision dynamics~\cite{MalevanetsJCP1999,gompperBookChap2009} (MPCD) and molecular dynamics (MD) (see Sec.~\ref{Sec:modeldetails}  for details of the simulation method~\cite{TheersSM2016,schwarzendahl2018maximum}). The MPCD part of the simulation method solves the Navier-Stokes equations to obtain the hydrodynamic interactions between rollers, while the MD part of the simulation takes care of the rigid body dynamics and steric interactions. The rollers are modeled as three dimensional dumbbells in a container with no-slip walls a the top and bottom and periodic boundary conditions otherwise. By means of a gravitational force the rollers sediment to the bottom plane. The Quincke-rolling is taken into account in an effective manner, by giving the roller an active torque that points along the rollers long axis in its body frame. Thereby, the roller effectively generates a hydrodynamic rotlet moment. 
Through the hydrodynamic coupling of the rotlet to the bottom wall the roller experiences a net flow, which in turn leads to a self-propulsion of the roller. The self-propulsion velocity increases with external torque (see Fig.~\ref{fig:spptorque}) and the simulation parameters were matched to obtain the experimental P\'eclet number $\mathcal{P} \approx 10^4$ (see also Sec.\ref{Sec:matching}).

First two dumbbells that initially rest on the bottom plane and are in a configuration akin to Fig~\ref{fig:fig2}(a) are simulated. The tetramer was stable until the end of the simulation time ($5\times 10^4 \tau_{\mathrm{MPCD}}$), while performing a total of $27$ rotations around each other. Hence, the hydrodynamic interactions between two dumbbells are also not enough to explain the break up of tetramers observed in experiments. Here, $\tau_{\mathrm{MPCD}}$ is the time unit of the MPCD simulation method (see also Sec.~\ref{Sec:compDetails}).

Second, a third dumbbell that passes by the tetramer was included into the simulation. Here, the tetramer breaks up due to long ranged hydrodynamic interactions mediated by the roller passing by (see Fig~\ref{fig:fig4}(a)), a situation that is typically observed in our 
experiments 
(see Fig~\ref{fig:fig4}(b)). 
In total $16$ passing by events (as in Fig~\ref{fig:fig4}(a)) were simulated, where $12$ times the tetramer broke up due to long ranged hydrodynamic interaction from the third dumbbell passing by. In the $4$ other cases all three dumbbells collide and break up due to steric interactions.
Mostly, (in $8$ out of $12$ cases) the long ranged hydrodynamic interactions are strong enough to break up the tetramer after the other dumbbell passes by for the first time. Here, the tetramer is able to perform half of a rotation around itself. In all other cases ($4$ out of $12$) the tetramer breaks up after the other dumbbell passes by for the second time (due to periodic boundary conditions), while the tetramer approximately performs one rotation around itself. 

\section{Conclusions}
In conclusion, by a combination of analytical theory and computer simulations, we have explored the stability of dumbbell-like active particles which interlock to form a jointly spinning pair. Our results show that for geometric interlocking stability is enhanced but depends on the activity. For weaker interlocking governed by soft monomeric interactions  stability is always lost.  Our results were compared to experimental data on dumbbell-like Quincke rollers. It was also shown that the stability can be lost due to hydrodynamic interactions of approaching neighboring particles.   
For the future, more complex particle shapes such as snowman~\cite{chaturvedi2012simple,kang2018influence}, colloidal bananas~\cite{ulbrich2022effect}, colloidal ``dolls''~\cite{kraft2013brownian} or mickey-mouse~\cite{wolters2015self} particles should be considered which provide more complex ways of activity-induced interlocking. The dependence on the direction of the self-propulsion relative to the particle orientation is another degree of freedom which should be analyzed and optimized. Also the most probable path of the dumbbells taken along the destabilization process should be analyzed by simulation, theory and experiment following recent ideas~\cite{zanovello2021optimal,zampetaki2021collective,yasuda2022most}.
Finally clusters with more than two particles should be analyzed with a systematic stability analysis which could provide a systematic understanding of the initial stages of motility-induced phase separation.

\begin{acknowledgments}
AMA thanks the funding support provided by CONACyT.
CPR wishes to acknowledge the Alexander von Humboldt foundation for a Bessel Award.

\end{acknowledgments}

\appendix

\section{Simulation with hydrodynamic interactions}
\label{Sec:modeldetails}
In order to simulate the hydrodynamic interactions between dumbbells a hybrid simulation scheme using a combination Molecular dynamics (MD) and Multiparticle collision dynamics (MPCD) is used. 
Each dumbbell is considered as a ridig body with center of mass position $\bm{R}$, mass $M$, and orientation $\bm{q}$, where $\bm{q}$ are quaternions (see below for a detailed explanation). 
A similar simulation method to the one presented here has been used in~\cite{TheersSM2016,schwarzendahl2018maximum}.

\subsection{Rigid body dynamics of rollers}
\label{Sec:rigidbodydyn}
The rigid-body dynamics of the rollers are determined by Newtons equations and equations for the roller's rotational degrees of freedom. Typically, the rotational degrees of freedom are parameterized by Euler angles, however, these are numerically unstable. Instead it is convenient to use another representation of the Euler angles (which is mathematically equivalent), which is given by quaternions
$\bm{q}= (q_0,q_1,q_2,q_3)^\mathrm{T}$, where
 $\mathrm{T}$ matrix transposition (see also~\cite{Allen1987}). 
The equations of equations of motion for a rigid body then read~\cite{OmelyanPRE1998}
\begin{align}
        m \ddot{\bm{R}} &= \bm{F} - g M \bm{e}_z\,,
        \label{eq:Newton2}
        \\
        \ddot{\bm{q}} &= \frac{1}{2} \left[ \mathbf{W}\left( \dot{\bm{q}}\right) 
        \begin{pmatrix}
                0\\ \bm{\varOmega}^\mathrm{b}
        \end{pmatrix}
        +
          \mathbf{W}\left( \bm{q} \right) 
        \begin{pmatrix}
        0\\  
        \dot{\bm{\varOmega}}^\mathrm{b}
        \end{pmatrix}
        \right]\,, \\
        \dot{ \bm{q}} &= \frac{1}{2} \mathbf{W}(\bm{q}) 
        \begin{pmatrix}
                0\\ \bm{\varOmega}^\mathrm{b}
        \end{pmatrix}\,, \\
        \dot{\Omega}_{\alpha}^\mathrm{b} &= (I_\mathrm{m}^\mathrm{b})^{-1}_{\alpha} \left( T^\mathrm{b}_{\alpha} + \left( (I_\mathrm{m}^\mathrm{b})_{\beta} - (I_\mathrm{m}^\mathrm{b})_{\gamma} \right) \Omega_{\beta}^\mathrm{b} \Omega_{\gamma}^\mathrm{b} \right).
\label{eq:RigidBodyAngle}
\end{align}
Here, $\mathbf{I}_\mathrm{m}^\mathrm{b}$ is the moment of inertia tensor of the roller in the body frame and
 $\bm{\varOmega}$  angular velocity of the roller. The
 indices $(\alpha,\beta,\gamma)$ take values of the cyclic permutations of $(x,y,z)$. In Eq.~\eqref{eq:Newton2} the steric forces are calculated from a potentials $\Phi_{pp}$, for particle-particle interactions and $\Phi_{\mathrm{wall}}$ for wall interactions  with
$\bm{F}=-\nabla (\Phi_{pp} +\Phi_{\mathrm{wall}})$ and $\bm{T}$  is the torque acting on a dumbbell. Further, a gravitational force acts on the dumbbells center of mass (second term on the right hand side of Eq.~\eqref{eq:Newton2}), which has a strength $g$ and points into the negative $z$-direction denoted by $-\bm{e}_z$.
The forces and torques that stem from steric interactions between rollers are mediated by a Weeks--Chandler--Andersen (WCA) potential \cite{WeeksJCP1971}, explicitly
\begin{equation} \label{eq:WCAPotential}
  \Phi_{pp} (r_{ij,ab}) = 
            4 \epsilon \left[ \left(\frac{\sigma_{ab}}{r_{ij,ab}} \right)^{12}-  
                              \left(\frac{\sigma_{ab}}{r_{ij,ab}} \right)^6 \right] 
            + \epsilon           
\end{equation}
if $r_{ij,ab} < 2^{1/6}\sigma_{ab}$, and $\Phi (r_{ij,ab}) = 0$ otherwise, 
where $\sigma_{ab}$ is the sum of the radii of sphere $a$ and sphere $b$, $\epsilon$ is the energy scale and
$r_{ij,ab}\equiv |\bm{r}_{ia}-\bm{r}_{jb}|$ is the distance between sphere $a$ of roller $i$ and sphere $b$ of roller $j$. The steric interactions with the walls at the top and bottom of the simulation box are also calculated through a WCA potential, which reads
\begin{equation} \label{eq:WCAPotentialwall}
  \Phi_{\mathrm{wall}} (z_i) = 
            4 \epsilon \left[ \left(\frac{\sigma_{a}}{|z_i-z_{\mathrm{virt}}| } \right)^{12}-  
                              \left(\frac{\sigma_{a}}{|z_i-z_{\mathrm{virt}}|} \right)^6 \right] 
            + \epsilon           
\end{equation}
if $|z_i-z_{\mathrm{virt}}| < 2^{1/6}\sigma_{a}$, and $\Phi (r_{ij,ab}) = 0$ otherwise, where $z_{\mathrm{virt}}= z_{\mathrm{bottom}} - \sigma_{a}/2 $ for the bottom wall and $z_{\mathrm{virt}}= z_{\mathrm{top}} + \sigma_{a}/2 $ for the top wall. Here, $z_{\mathrm{top}}$ and $z_{\mathrm{bottom}}$ are the positions of the top and bottom walls respectively and $\sigma_a$ is the diameter of sphere $a$.

In the simulation, vectors need to be transformed between the body frame and the laboratory frame. A vector in the laboratory frame $\bm{f}$ which is transformed to the body frame  vector $\bm{f}^\mathrm{b}$ is given by
\begin{align}
        \bm{f}^\mathrm{b} = \mathbf{D} \bm{f},
        \label{eq:BodyLabTrafo}
\end{align}
where the matrix $\mathbf{D}(\bm{q})$ is constructed from the quaternions, explicitly (see also \cite{Allen1987})
\begin{align}
        &D = \nonumber \\
       &\begin{pmatrix}
    q_0^2 + q_1^2 -q_2^2 - q_3^2   &    2\left( q_1 q_2 + q_0 q_3 \right)    &     2\left( q_1 q_2 - q_0 q_2 \right) \\
    2\left( q_2 q_1 -q_0 q_3 \right)       &    q_0^2 - q_1^2 + q_2^2 - q_3^2    &    2\left( q_2 q_3 + q_0 q_1 \right) \\
    2\left( 2 q_3 q_1 + q_0 q_2  \right)    &    2\left( q_3 q_2 - q_0 q_1 \right)    &  q_0^2 - q_1^2 - q_2^2 + q_3^2 \\
 \end{pmatrix} .
        \label{eq:LabBodyTrafo}
\end{align}
The orientation of the roller can then be found by $\mathbf{D}^{-1}(\bm{q}(t))(0, 0, 1)^\mathrm{T}$.
In the following, all quantities with an index $b$ are computed in the body frame, all other quantities are in the laboratory frame.
Further, the matrix $\mathbf{W}$ is given by (see also \cite{Allen1987})
\begin{align}
        \mathbf{W}\left( \bm{q} \right) = 
       \begin{pmatrix}
  q_0 & -q_1 & -q_2 & -q_3 \\
  q_1 & q_0 & -q_3 & q_2 \\
  q_2 & q_3 & q_0 & -q_1 \\
  q_3 & -q_2 & q_1 & q_0 \\
 \end{pmatrix} .
        \label{eq:Qmatrix}
\end{align}

The rotation each roller, mediated by the Quincke effect, is included in an effective manner by a constant torque acting on each roller. This torque is chosen to be along the $z$-axis in the rollers body frame. 
Hence, the torque in the body frame $\bm{T}^b=\bm{T}^b_{\mathrm{steric}} + \bm{T}^b_{\mathrm{active}}$ has two contributions, the torque stemming from steric interactions $\bm{T}^b_{\mathrm{steric}}= \bm{D} \bm{R}_F \times \bm{F}$ and the torque stemming from the Quincke rotation  $\bm{T}^b_{\mathrm{active}}= (0,0,T^b_{\mathrm{active}})$.  Here, the vector  $\bm{R}_\mathrm{F}$ is connecting the center of mass of the 
roller to the point of contact with the neighbor or wall.

To intergrate Eq.~\eqref{eq:Newton2}-\eqref{eq:RigidBodyAngle} a Verlet algorithm \cite{OmelyanPRE1998} is used (see also \cite{TheersSM2016, schwarzendahl2018maximum}).

\subsection{Center of mass and moment of inertia}
\label{subsec:CoM}
The roller has two spheres which are slighly unequal in size due to experimental limitations. In the simulations this asymmetry is taken into account and hence sphere $a$ has a diameter $\sigma_a$ and sphere $b$ a diameter $\sigma_b$. The center of mass then changes because of the asymmetry and is shifted into the larger sphere.
In the body frame, the roller is aligned with the $z$ direction and the coordinates of the centers of the $a$ and $b$ spheres are $z_a$ and $z_b$, respectively. The $z$-component of the center of mass of the roller is then given by
\begin{align}
z_{\text{CoM}} = \frac{ V_a z_a + V_b z_b  
        }   { V_a + V_b },
    \label{eq:CenterOfMass}
\end{align}
 $V_i$ are the volumes of the respective spheres.

Further, the moment of inertia of the dumbbell is needed in the simulations. The moment of inertia of a sphere is given by $I_{Sp_i}= \frac{8}{15} \rho_s \pi (d_i/2)^5 $, where $\rho_s$ is the spheres density and $d_i$ its diameter. By using the parallel axis theorem the moments of inertia of the roller are
\begin{align}
        &I_{(x,y)} = I_{Sp_1, (x,y)}+ \rho \left( V_1  \right) x_1^2
        \label{eq:MomentOfInertiaxy}
        \\
        &I_z = I_{Sp_1,z} + I_{Sp_2,z}.
        \label{eq:MomentOfInertiaz}
\end{align}

\subsection{Multiparticle collision dynamics}
\label{subsec:MPCalgorithm}
To simulate the hydrodynamic interactions between roller the Multiparticle collision dynamics~\cite{MalevanetsJCP1999} technique, which is a mesoscale method that solves the Navier--Stokes equations is used. 
The simulated fluid has a density $\rho$ and temperature $T$. An Andersen thermostat and angular momentum conservation is used in the simulation, the specitic algorithm is called MPC-AT+a~\cite{NoguchiEPL2007,GotzePRE2007,gompperBookChap2009}.

In MPCD $N_\mathrm{fl}$ point-like particles of mass $m$ that perform a streaming step and a collision step, are used as an effective representation of the fluid. The particles have positions $\bm{r}_i \,,\, i\in [1,N_\mathrm{fl}] $ and during the streaming step are advected by
\begin{equation}\label{eq:MPCaStream}
        \bm{r}_i (t + \delta t) =  \bm{r}_i (t) +  \bm{v}_i(t) \delta t,
\end{equation}
where $\bm{v}_i(t)$ are the particles velocities and $\delta t $ is the MPCD timestep. 

During the collision step the fluctuating part of the particle's velocity is randomized, which effectively models interactions with other particles.
The collsion step is performed on a grid with $N_c$ collision cells and a lattice constant $a$. 
The set of particles that are in the same cell as particle $i$ is denoted by $\mathsf{C}(i)$.
The center of mass velocity in the cell $\mathsf{C}(i)$ is kept constant while the 
flucutating part randomized, explicitly each particle velocity is updated by
\cite{NoguchiEPL2007}
\begin{align}
        \bm{v}_i' = & \frac{1}{N_{\mathsf{C}(i)}} \sum_{j \in \mathsf{C}(i)} \bm{v}_{j}  +\bm{v}_i^{\text{ran}} - \frac{1}{N_{\mathsf{C}(i)}}\sum_{j \in \mathsf{C}(i)} \bm{v}_{j}^{\text{ran}}  \nonumber  \\
        &+ m  \left\lbrace\bm{\Pi}^{-1} \sum_{j \in \mathsf{C}(i)}\left[ \bm{r}_{j,c} \times \left( \bm{v}_i - \bm{v}_i^{\text{ran}}\right) \right] \right\rbrace\times \bm{r}_{i,c},
        \label{eq:MPCaCollision}
\end{align}
where $\bm{v}_i^{\text{ran}}$ is a random velocity drawn from Gaussian distribution with correlations $\sqrt{k_\mathrm{B}T/m}$. Here, the number of fluid and ghost particles (see Sec.~\ref{subsec:coupling}) in cell $\mathsf{C}(i)$ is given by $N_{\mathsf{C}(i)}$. Further,
$\bm{\Pi}^{-1}$ is the inverse of the moment of inertia tensor $ \bm{\Pi}\equiv \sum_{j \in \mathsf{C}(i)}m\left[(\bm{r}_j\cdot\bm{r}_j){\mathbf{I}}-\bm{r}_j\otimes\bm{r}_j\right]$ for the fluid particles in cell $\mathsf{C}(i)$ and $\bm{r}_{j,c}$ is the position of particle $j$ relative
to the center of mass of the cell $\mathsf{C}(i)$. 

Finally, to obtain Galilean invariance, a grid shift~\cite{IhlePRE2001} in $[-a/2, a/2]$ is performed after every streaming step.

\subsection{Coupling of the roller's and fluid's dynamics}\label{subsec:coupling}
On the roller's surface and on the top an bottom boundary of the simulation box we apply no-slip boundary conditions for the fluid.
\subsubsection{Streaming step}
During the streaming step the 
bounce-back rule~\cite{LamuraEPL2001} is applied. Hence, when a particle penetrates into a surface (particle or wall) during the streaming step, it is propagated back the same distance is traveled within the particle or wall. Further, its velocity is reversed. Note that multiple collisions~\cite{PaddingPRE2006} are possible and taken into account. 

From the collision between fluid particle and roller there will be a momentum transfer, which reads 
\begin{align}
        \bm{J}_i = 2 m \left[ \bm{v}_i - \bm{U} - \bm{\varOmega} \times \left( \bm{{\tilde{r}}}_i - \bm{R} \right)\right].
        \label{eq:BounceBackLinMomentumChange}
\end{align}
Here, $\bm{U}$ is the roller velocity, $\bm{\varOmega}$ is its angular velocity and 
$\bm{{\tilde{r}}}_i$ is the position of the fluid particle upon collision.
The fluid velocity then updates to
\begin{align}
        \bm{v}'_i= \bm{v}_i - \bm{J}_i/m.
        \label{eq:MPCpartBounceBack}
\end{align}
Further, the linear and angular velocity of the rollers are updated to
\begin{align}
        \bm{U}' &= \bm{U} + \sum_i \bm{J}_i /M
        \\
        \bm{\varOmega}' &= \bm{\varOmega} + \mathbf{I}_m^{-1} \sum_i (\bm{r}_i - \bm{R}) \times \bm{J}_i\,.
        \label{eq:rollerBounceBack}
\end{align}

\subsubsection{Collision step}
It has been shown that placing 
ghost particles~\cite{GotzePRE2007} inside walls are needed to obtain a no-slip boundary condition. Therefore, ghost particles $\bm{r}_i^g$ uniformly 
distributed within the roller and below the walls. The density of ghost particles within a roller and below the wall is matched to the fluids density. 

The positions of the ghost particles within the walls are kept constant and their velocity is randomized before each collision step, where we velocities components are drawn from a Gaussian with zero mean and correlations $\sqrt{k_\mathrm{B}T/m}$.

The ghost particles which are inside of the rollers with velocities $\bm{v}_i^g$ are updated before each collision step according to
\begin{align}
        \bm{v}_i^g = \bm{U}  + \bm{\varOmega} \times \left( \bm{r}_i^g - \bm{R} \right) + \bm{v}_i^{\text{ran}},
        \label{eq:GhostInitialv}
\end{align}
 where again $\bm{v}_i^{\text{ran}}$ are sampled from a Gaussian distribution. Afterwards the ghost particles take part collision step according to Eq.~\eqref{eq:MPCaCollision}.

As a result of the collision step, there will be a momentum $\bm{J}_i^g = m\left( \bm{v}_i^{g\prime} -\bm{v}_i^g \right)$ and angular momentum transfer $\bm{L}_i^g = \left( \bm{r}_i^g - \bm{R}  \right) \times \bm{J}_i^g$. 
Therefore the rollers~\cite{GotzePRE2010} velocity and angular velocity are updated to
\begin{align}
         \bm{U}' &= \bm{U} + \sum_i \bm{J}_i^g /M\,,
 \label{eq:rollerGhostMomentumChange}
        \\
        \bm{\varOmega}' &= \bm{\varOmega} + \mathbf{I}_\mathrm{m}^{-1} \sum_i \bm{L}_i^g\,.
        \label{eq:rollerGhostAngularMomentumChange}
\end{align}

\subsection{Computational details}
\label{Sec:compDetails}
We performed simulations in a three dimensional box with wall at the top and bottom of the simulation box and periodic boundary condition in the two other directions. The simulation container has a size of $50a \times 50a \times 50a$, where the walls where placed at $z_{\mathrm{bottom}}= a$ and  $z_{\mathrm{top}}= 49a$, leaving space for ghost particle below the walls.
A total of $\langle N_\mathsf{C} \rangle = 20$ fluid particles were used per cell and the MPCD timestep was chosen as $\delta t = 10^{-2} \tau_{\mathrm{MPCD}}$, 
whereas the MD timestep is $\delta t_{\mathrm{MD}} =10^{-4} \tau_{\mathrm{MPCD}}$, where $\tau_{\mathrm{MPCD}}= \sqrt{m a^2/(k_{\mathrm{B}} T)}$ is the unit of time in the MPCD simulation. 
The kinematic viscosity $\nu=\eta/\rho$ of the fluid for the MPC-AT+a algorithm (including both kinetic and collisional contribution) is then given by
$\nu = 3.88 a \sqrt{k_\mathrm{B} T/m}$~\cite{GotzePRE2007,noguchiPRE2008,gompperBookChap2009}. The energy scale for steric interaction was chosen as 
$\epsilon= 200 k_{\mathrm{B}} T$.

The size of the rollers spheres was chosen slightly unequal to resemble the experiments with $\sigma_a= 6a$ and $\sigma_b= 5.5a$. The strength of the gravitational force is chosen as $g= 50 k_{\mathrm{B}} T/a^2$ to ensure that the rollers sediment to the bottom wall.

\subsection{Matching parameters to experiments}
\label{Sec:matching}
\begin{figure}
    \centering
    \includegraphics[width=0.5\textwidth]{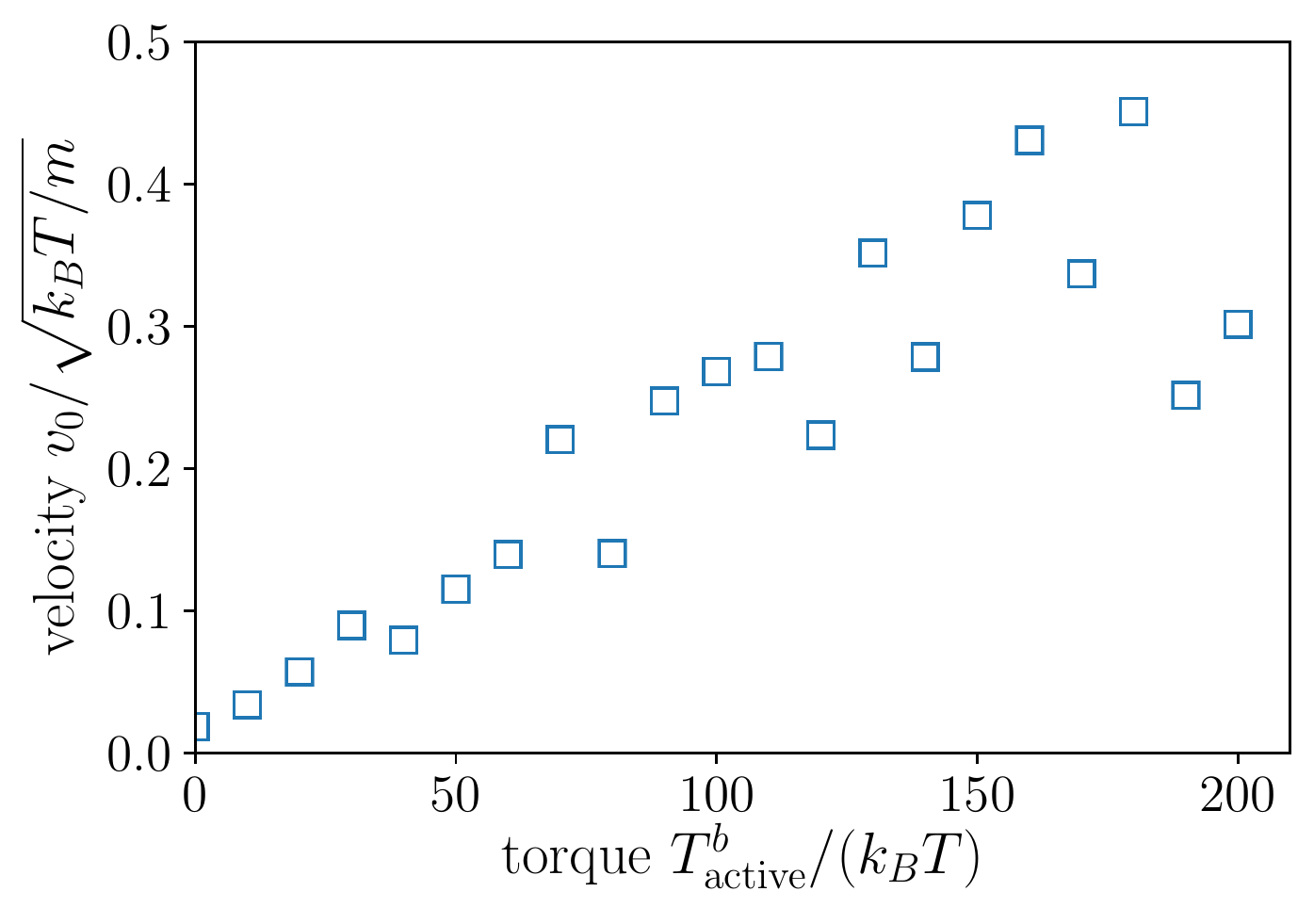}
    \caption{Self-propulsion velocity as a function of active torque.}
    \label{fig:spptorque}
\end{figure}
The experimentally measured P\'eclet number is $\mathcal{P} \approx 10^4$. To match this number in the simulations, first the translational diffusion coefficient of a dumbbell without actice  torque $T^b_{\mathrm{active}}$ was measured, resluting in $D^t= 2.3 \times 10^{-5} \sqrt{a^2 k_{\mathrm{B}} T /m}$. Using the definition of the P\'eclet number $\mathcal{P}= v_0 \sigma/D^t$, the self propulsion needed in the simulations to match experiments is $v_0 = 0.26 \sqrt{k_B T/m}$. 

Figure~\ref{fig:spptorque} shows the self-propulsion of an isolated roller as a function of active torque. Using this graph we find that an active torque $T^b_{\mathrm{active}} \approx 120 k_B T$ yields the correct self-propulsion velocity to match the experiments.



\providecommand{\noopsort}[1]{}\providecommand{\singleletter}[1]{#1}%

\end{document}